\documentclass[amsmath,amssymb]{revtex4}

\usepackage{graphicx}
\usepackage{dcolumn}
\usepackage{bm}

\begin{document}

\title{A note on the calculation of the effective range}

\author{H. Ouerdane and M. J. Jamieson}

\address{Department of Computing Science, University of Glasgow, Glasgow G12 8QQ, UK.}

\begin{abstract}
The closed form of the first order non-linear differential equation that is satisfied by the effective range within the variable phase formulation of scattering theory is discussed. It is shown that the conventional method of determining the effective range, by fitting a numerical solution of the Schr\"{o}dinger equation to known asymptotic boundary conditions, can be modified to include the first order contribution of a long range interaction.
\end{abstract}


\maketitle

\section{Introduction}
Scattering of atoms and molecules in an ultra-cold environment is a process that is very important in the interpretation of measurements of properties of cold trapped gases and in the study of Bose-Einstein condensation. At such low temperatures s-wave scattering dominates and the scattering length, $a$, usually suffices to describe collisions. However the effective range, $r_{\rm e}$, is important because it implies that there is a relative contribution to the elastic scattering cross section at energy $E$ of $2 \mu Ear_{\rm e}$ where $\mu$ is the reduced mass. The effective range can provide an upper bound to the temperature, $T$, of a trapped sample in thermal equilibrium below which only the scattering length is significant; for example the corresponding relative contribution to the elastic scattering cross section that is attributable to the effective range is $2 \mu k_{\rm B}Tar_{\rm e}$ where $k_{\rm B}$ is Boltzmann's constant. Previously \cite{r1} we used the  variable phase formulation of scattering theory \cite{r2} to study the calculation of the scattering length in the presence of long range forces. Here we make a study of the effective range.

In the conventional method of calculating the effective range a numerical solution of the Schr\"{o}dinger equation for scattering is fitted to known asymptotic boundary conditions at some suitably large separation $R_{\rm c}$ to yield a phase shift. This is repeated for several small values of the asymptotic wave number of the relative motion, $k$, and the values of $a$ and $r_{\rm e}$ are obtained from the effective range theory fit of the phase shift as a function of $k$ \cite{r3}. This procedure is appropriate even in the presence of long range forces because, by making the fit at the finite separation $R_{\rm c}$, we solve a problem for short ranges forces curtailed at $R_{\rm c}$. A difficulty occurs in that $R_{\rm c}$ must be large. Here we present a correction that accounts for the influence of a long range force over the range $[R_{\rm c}, \infty[$ to first order in the potential strength. It allows much smaller values of $R_{\rm c}$ to be used with attendant reduction in computation time and accumulated error. Marinescu \cite{r4} has provided first and higher order corrections but they are complicated in comparison to our formulation and, in most cases, a first order correction is enough.

\section{Theory}

In the notation of Levy and Keller \cite{r5} the s-wave phase shift $\eta$ is expanded as

\begin{equation}
\eta=N \pi + k(\eta_0 + k^2 \eta_1 + \ldots)
\label{eqn1}
\end{equation}

where $N$ is the number of bound states supported by the interaction potential. Equation (\ref{eqn1}) is in accord with Levinson's theorem. Levy and Keller \cite{r5} omitted the term $N \pi$ from their equation (8) but the subsequent analysis is not altered because effective range theory \cite{r3} provides an expansion for $k \cot \eta$. 

When long range forces are present expansion (\ref{eqn1}) and the effective range expansion have additional terms in $k$, such as $k^4\ln k$, but for interactions dominated by a Van der Waals $R^{-6}$ dispersion potential the leading terms have the indicated $k$-dependence although the coefficients, $\eta_i$, are modified \cite{r6}. In variable phase theory the phase shift and these coefficients are functions of the separation, $R$, and hence the scattering length and the effective range are also functions of $R$. The effective range expansion is \cite{r3}

\begin{equation}
k \cot \eta(R)=-\frac{1}{a(R)} + \frac{1}{2} r_e(R) k^2 + \ldots .
\label{eqn1a}
\end{equation}

The scattering length, $a(R)$, is the limit as $k \rightarrow 0 $ of $-\tan \eta(R)/k$ and hence is $-\eta_0(R)$. The effective range is twice the coefficient of the term in $k^2$ in equation (\ref{eqn1a}) and from the Taylor expansion of the cotangent it can be expressed as 

\begin{equation}
r_{\rm e}(R)=\frac{2a(R)}{3}-\frac{2\eta_1(R)}{[a(R)]^2}.
\label{eqn2}
\end{equation}

The usual scattering length and effective range are the asymptotic values taken by $a(R)$ and $r_{\rm e}(R)$ as $R \rightarrow \infty$.

The log-derivative in variable phase theory is 

\begin{equation}
u(R)=\left [ k+\frac{{\rm d} \eta(R)}{{\rm d} R} \right ] \cot[kR+ \eta(R)].
\label{eqn2a}
\end{equation}

Substitution of $u(R)$ into the Riccati equation that is satisfied by the log-derivative yields the variable phase equation (equation (6) of Levy and Keller \cite{r5})

\begin{equation}
\frac{{\rm d} \eta(R)}{{\rm d} R}=-\frac{1}{k}V(R) \sin^2[kR+\eta(R)]  
\label{eqn2b}
\end{equation}

\noindent where $V(R)$ denotes $2 \mu/\hbar^2$ times the interaction potential. From equation (\ref{eqn1}) Levy and Keller \cite{r5} rearranged equation (\ref{eqn2b}) as a power series in $k$ to obtain non-linear first order differential equations in $R$ that are satisfied by the coefficients $\eta_i(R)$ (their equations (9) and (10)) from which we find

\begin{equation}
\frac{{\rm d} a(R)}{{\rm d} R}=V(R)[R-a(R)]^2  
\label{eqn3}
\end{equation}

\noindent and

\begin{equation}
\frac{{\rm d} \eta_1(R)}{{\rm d} R}=-2V(R)[R-a(R)]\eta_1(R) + \frac{1}{3}V(R)[R-a(R)]^4.  
\label{eqn4}
\end{equation}

\subsection{Numerical procedures}

Although equation (\ref{eqn4}) has the closed form solution,

\begin{equation}
\eta_1(R) = \frac{2}{3} \exp\left\{-2 \int^R V(S)[S-a(S)] {\rm d}S\right\}
\times
\int^R\exp\left\{2 \int^S V(Q)[Q-a(Q)] {\rm d}Q \right\}V(S)[S-a(S)]^4{\rm d}S,
\label{eqn5}
\end{equation}

\noindent we find that numerical evaluation is not conveniently provided in this way because the closed form solution (\ref{eqn5}) requires computation of a double integral and, when the interaction supports bound states, also requires a suitable numerical account of the poles in the scattering length, $a(R)$; these poles also make a direct numerical solution of equation (\ref{eqn4}) impractical. The adaptive method of Lambert \cite{r7} is unsuitable for heavy particle collisions because there are many poles in $a(R)$. The log-derivative method that we used previously \cite{r1} is also unsuitable; although equations (\ref{eqn3}) and (\ref{eqn4}) may be written in the Riccati form

\begin{equation}
\frac{{\rm d}}{{\rm d} R}
\left[ \begin{array}{cc}
R-a(R) & \eta_1(R) \\
0      & R-a(R)
\end{array} \right]
+ V(R)
\left[ \begin{array}{cc}
R-a(R) & \eta_1(R) \\
0      & R-a(R)
\end{array} \right]^2
=
\left[ \begin{array}{cc}
1 & \frac{\displaystyle 1}{\displaystyle 3}V(R)[R-a(R)]^4 \\
0 & 1
\end{array} \right]
\label{eqn6}
\end{equation}

\noindent the difference between the right hand side and the unit matrix prevents application of the log-derivative method. A rational fraction transformation of equation (\ref{eqn6}) yields another Riccati equation \cite{r7a} but there is no obvious transformation that simultaneously changes the right hand side of equation (\ref{eqn6}) to a unit matrix and maintains the structure of its left hand side.

We suggest that the conventional method for obtaining the effective range is used at separation $R_{\rm c}$, chosen beyond the last pole in $a(R)$, and equations (\ref{eqn2}) and (\ref{eqn4}) are then used to obtain a correction that is of first order in the potential strength.

\section{Long range corrections} 
The exponential terms in expression (\ref{eqn5}) are positive. Hence the contribution to $\eta_1(R)$ from the range $[R_{\rm c}, \infty)$ is negative when the potential is attractive over this range and therefore the conventional calculation yields an upper bound to $\eta_1=\eta_1(\infty)$.

The subsequent analysis is simpler when presented in terms of dimensionless quantities obtained by scaling by suitable powers of $R_{\rm c}$. Throughout the remainder of this note, unless otherwise indicated, we let $R$, $a$, $r_{\rm e}$, $V$ and $\eta_1$ denote the corresponding quantities divided by $R_{\rm c}$, $R_{\rm c}$, $R_{\rm c}$, $R_{\rm c}^{-2}$ and $R_{\rm c}^3$ respectively. Equations (\ref{eqn2}), (\ref{eqn3}) and (\ref{eqn4}) are not altered. The correction, $\delta \eta_1$, to $\eta_1$, obtained by integrating the right hand side of equation (\ref{eqn4}) by parts over the range $[1,\infty[$, is

\begin{equation}
\delta \eta_1=-2 \eta_1 \left[V^{(1)}(a-1)+V^{(2)}\right]-
\frac{1}{3}\sum_{i=1}^5 \frac{4!}{(5-i)!}V^{(i)}(a-1)^{5-i}
\label{eqn7}
\end{equation}

\noindent where 

\begin{equation}
V^{(i)}(R)=\int^R V^{(i-1)}(S){\rm d}S    
\label{eqn8}
\end{equation}

\noindent with $i=1,2,\ldots,5$ and $V^{(0)}(R)$ identified as $V(R)$, and the right hand expression is evaluated at $R=1$ ({\it i.e.} at separation $R_{\rm c})$.

The inverse power potential that is given in terms of the unscaled separation by $-C_n R^{-n}$ is represented in the scaled quantities by

\begin{equation}
V(R)=-\frac{\alpha_n^{n-2}}{R^n}
\label{eqn9}
\end{equation}

\noindent where

\begin{equation}
\alpha_n=\left(\frac{2 \mu C_n}{\hbar^2}\right)^{\frac{1}{n-2}}\frac{1}{R_{\rm c}}.
\label{eqn10}
\end{equation}

From equations (\ref{eqn2}), (\ref{eqn3}),  (\ref{eqn7}), (\ref{eqn8}), (\ref{eqn9}) and (\ref{eqn10}) we obtain the first order correction, $\delta r_{\rm e}$, to the effective range as 

\begin{equation}
\delta r_{\rm e}=\alpha_n^{n-2}\left[\frac{2 r_{\rm e}}{(n-3)a}-\frac{2 r_{\rm e}}{(n-2)}+\frac{2}{3(n-5)a^2}-\frac{8}{3(n-4)a}+\frac{2}{n-3}\right]\label{eqn11}
\end{equation}

\noindent where the right hand side is evaluated at $R=1$. Equation (\ref{eqn11})is consistent with the correction that was derived by different methods \cite{r9}. Note that the correction given by equation (\ref{eqn11}) is not {\em identical} to the correction in reference \cite{r9}; equation (\ref{eqn11}) provides the correction to the effective range whereas reference \cite{r9} gives the correction to the product $r_{\rm e}a^2$. Their equivalence is readily shown by introducing the correction to the scattering length; the correction is \cite{r9}

\begin{equation}
\delta a=-\alpha_n^{n-2}\left[\frac{1}{n-3}-\frac{2a}{n-2}+\frac{a^2}{n-1}\right].
\label{eqn11a}
\end{equation}

\section{Method}
Our suggested procedure for calculating the effective range is as follows. First evaluate the scattering length and effective range at separation $R_{\rm c}$ by the conventional method and then scale them by division by $R_{\rm c}$. For each term such as (\ref{eqn9}) of the long range potential, evaluate the scaled correction $\delta r_{\rm e}$ of equation (\ref{eqn11}) and find the sum of the corrections; this is legitimate because we are interested in only the first order correction. Multiply the resulting correction by $R_{\rm c}$. Some experiment is needed to find a suitable value for $R_{\rm c}$ but generally it is much smaller than the separation that represents infinity in the conventional method without correction. The leading terms that describe the influence of the long range force that are not accounted for by equation (\ref{eqn11}) are of second order and hence are proportional to $(\alpha_n^{n-2})^2$. The chosen value of $R_{\rm c}$ should coerce this quantity to be smaller than unity; a suitable choice is several times the characteristic length, $\left(2 \mu C_n/\hbar^2\right)^{1/n-2}$, of the potential.

\section{Example}
For the model Cs-Cs potential described by Gribakin and Flambaum \cite{r8} we obtained the effective range accurate to three significant figures with $R_{\rm c}=1400$ bohr while even the first significant figure is uncertain in the conventional uncorrected calculation with $R_{\rm c}=10000$ bohr; the characteristic length of the dominant term of the asymptotic dispersion potential is 203 bohr. We show details of the convergence in table \ref{Table1}; the values shown in table \ref{Table1} are unscaled. The characteristic lengths of the dispersion terms in $C_6$, $C_8$ and $C_{10}$ are, in bohr, 203, 80 and 50 respectively.

\begin{table*}
\caption{Effective Range and Corrections (bohr)\label{Table1}}
\begin{ruledtabular}
\begin{tabular}{ccccccc}
Separation $R_{\rm c}$ & Effective Range & 
\multicolumn{3}{c}{Correction from term in} & Effective Range \\
(bohr) & & $C_6$ & $C_8$ & $C_{10}$ & (Corrected) \\
\hline
 600 & 284 & 317 & 428 $\times 10^{-4}$ & 1088 $\times 10^{-8}$ & 602 \\
 800 & 358 & 259 & 200 $\times 10^{-4}$ &  287 $\times 10^{-8}$ & 617 \\
1000 & 406 & 215 & 107 $\times 10^{-4}$ &   99 $\times 10^{-8}$ & 621 \\
1200 & 440 & 183 &  64 $\times 10^{-4}$ &   41 $\times 10^{-8}$ & 623 \\
1400 & 464 & 160 &  41 $\times 10^{-4}$ &   19 $\times 10^{-8}$ & 624 \\
1600 & 483 & 141 &  28 $\times 10^{-4}$ &   10 $\times 10^{-8}$ & 624 \\
1800 & 498 & 126 &  20 $\times 10^{-4}$ &    6 $\times 10^{-8}$ & 624 \\
2000 & 510 & 115 &  15 $\times 10^{-4}$ &    3 $\times 10^{-8}$ & 624 \\
3000 & 546 &  78 &   4 $\times 10^{-4}$ &      $< 10^{-8}$    & 624   \\
4000 & 565 &  59 &   2 $\times 10^{-4}$ &      $< 10^{-8}$    & 624   \\
5000 & 576 &  47 &     $10^{-4}$        &      $< 10^{-8}$    & 624   \\
6000 & 584 &  40 &     $< 10^{-4}$      &      $< 10^{-8}$    & 624   \\
\end{tabular}
\end{ruledtabular}
\end{table*}

\section{Conclusion}
The variable phase method provides a simple way in which account may be taken of the long range dispersion terms of an interatomic potential in calculations of the effective range. The method gives considerable saving of computation time and of accumulated truncation error.

\section*{Acknowledgments}
   
We are pleased to thank Dr. D. Vrinceanu for discussions about the variable phase method. This work was supported by the Engineering and Physical Sciences Research Council. 

\bibliography{jphysb3}

\end{document}